\documentclass[prl,aps,superscriptaddress]{revtex4}
\usepackage{amsfonts}

\usepackage{dcolumn}
\usepackage{bm}
\usepackage{tikz}
\usepackage[colorlinks=true,citecolor=blue,urlcolor=blue]{hyperref}
\usepackage{amsmath,amsfonts,amssymb,times,natbib}
\usepackage[standard]{ntheorem}

\begin{document}
\noindent

\title{Bell's Nonlocality Can be Detected by the Violation of Einstein-Podolsky-Rosen Steering Inequality}

%\title{One-Way Einstein-Podolsky-Rosen Steering and Steering-Induced Bell's Nonlocality}

%\title{Experimental Proposals to Test One-Way Steering and Steering-Induced Bell's Nonlocality}

\author{Jing-Ling Chen\footnote{Correspondence to
J.L.C. (chenjl@nankai.edu.cn).}}
 \affiliation{Theoretical Physics Division, Chern Institute of Mathematics, Nankai University,
 Tianjin 300071, People's Republic of China}
 \affiliation{Centre for Quantum Technologies, National University of Singapore,
 3 Science Drive 2, Singapore 117543, Singapore}

\author{Changliang Ren\footnote{Correspondence to
C.R. (renchangliang@cigit.ac.cn).}}
 \affiliation{Center for Nanofabrication and System Integration, Chongqing Institute of Green and Intelligent Technology, Chinese Academy of Sciences, Chongqing 400714, People's Republic of China}

\author{Changbo Chen}
 \affiliation{Chongqing Key Laboratory of Automated Reasoning and Cognition, Chongqing Institute of Green and Intelligent Technology, Chinese Academy of Sciences, Chongqing 400714, People's Republic of China}

\author{Xiang-Jun Ye}
 \affiliation{Key Laboratory of Quantum Information, University of Science and Technology of China, University of Science and Technology of China, Hefei 230026, People's Republic of China}
\affiliation{Synergetic Innovation Center of Quantum Information and Quantum Physics, University of Science and Technology of China, Hefei 230026, People's Republic of China}

\author{Arun Kumar Pati\footnote{Correspondence to
A.K.P. (akpati@hri.res.in).}}
 \affiliation{Quantum Information and Computation Group, Harish-Chandra Research Institute, Chhatnag Road, Jhunsi, Allahabad 211019, India}

\date{\today}

\maketitle

%\emph{Introduction.}---

%\section{Introduction}

\textbf{Recently quantum nonlocality has been classified into three distinct types: quantum entanglement, Einstein-Podolsky-Rosen steering, and Bell's nonlocality. Among which, Bell's nonlocality is the strongest type. Bell's nonlocality for quantum states is usually detected by violation of some Bell's inequalities, such as Clause-Horne-Shimony-Holt inequality for two qubits. Steering is a manifestation of nonlocality intermediate between entanglement and Bell's nonlocality. This peculiar feature has led to a curious quantum phenomenon, the one-way Einstein-Podolsky-Rosen steering. The one-way steering was an important open question presented in 2007, and positively answered in 2014 by Bowles \emph{et al.}, who presented a simple class of one-way steerable states in a two-qubit system with at least thirteen projective measurements. The inspiring result for the first time theoretically confirms quantum nonlocality can be fundamentally asymmetric. Here, we propose another curious quantum phenomenon: Bell nonlocal states can be constructed from some steerable states.  This novel finding not only offers a distinctive way to study Bell's nonlocality without Bell's inequality but with steering inequality, but also may avoid locality loophole in Bell's tests and make Bell's nonlocality easier for demonstration. Furthermore, a nine-setting steering inequality has also been presented for developing more efficient one-way steering and detecting some Bell nonlocal states.}

\vspace{5mm}

In 1935, the famous Einstein, Podolsky and Rosen (EPR) paper
indicated that
%there were some conflicts between
quantum mechanics is in conflict with the notion of locality and reality~\cite{EPR}. If local realism is correct, then
quantum mechanics cannot be considered as a complete theory to describe physical reality. Immediately after the publication of the EPR paper, Schr\"odinger made a response by
conjuring two
important notions, namely, the quantum \emph{entanglement} and the quantum \emph{steering}. According to Schr\"odinger, quantum entanglement is
``the characteristic trait of quantum mechanics" that
distinguishes quantum theory from classical theory ~\cite{Schrodinger35}. The notion of ``steering" is closely related to the statement of ``spooky action at a distance",
which Einstein was disturbed all the time. EPR steering reflects such a ``spooky action" feature that manipulating one object  seemingly affects another instantaneously,
even it is far away.

Different to Schr\"odinger's response, in 1964, Bell proposed an inequality for local hidden variable (LHV) models ~\cite{Bell}.
%by investigating two entangled spin-1/2 systems (i.e., two qubits)~\cite{Bell}.
The violation of Bell's inequality by quantum entangled states implies Bell's nonlocality. This is well-known as Bell's theorem, which has
established what quantum theory can tell us about the fundamental features of \emph{Natur}e, and been widely regarded as ``the most profound discovery of science"~\cite{Stapp}.
Until now, the fundamental theorem has achieved ubiquitous applications in different quantum information tasks, such as quantum key distribution~\cite{crypt}, communication
complexity~\cite{Brukner}, and  random number generation~\cite{Random10}.

Unlike quantum entanglement and Bell's nonlocality, the research
field of quantum steering has been sterile till 2007, when Wiseman,
Jones, and Doherty~\cite{WJD07} reformulated the idea and placed it
firmly on a rigorous ground. Since then EPR steering has gained a
very rapid development in both theories
~\cite{SU,QKD,AVN,He,Jevtic,Skrzypczyk,Bowles,Piani} and experiments
~\cite{NP2010,NC,PRX,NJP, AS12,Schneeloch,Sun,Li,Wollmann,Sun2}.
Most research topics as well as research approaches in the field of
Bell's nonlocality have been transplanted similarly to the field of
EPR steering. For instance, steering inequalities have been proposed
to reveal the EPR steerability of quantum states, very similar to
the violation of Bell's inequalities reveals Bell's nonlocality.

According to Ref.~\cite{WJD07}, entanglement, EPR steering and
Bell's nonlocality are called by a joint name as `` quantum
nonlocality", which has an interesting hierarchical structure:
quantum entanglement is a superset of steering, and Bell's
nonlocality is a subset of steering. However, among the three types
of quantum nonlocality, only steering can possess a curious feature
of ``one-way quantumness". Suppose Alice and Bob share a pair of
two-qubit state, it is not hard to imagine that if Alice entangles
with Bob, then Bob must also entangle with Alice. Such a symmetric
feature holds for both entanglement and Bell nonlocality. However,
the situation is dramatically changed when one turns to a novel kind
of quantum nonlocality in the middle of entanglement and Bell
nonlocality, the EPR steering. It may happen that for some
asymmetric bipartite quantum states, Alice can steer Bob but Bob can
never steer Alice. This distinguished feature would be useful for
some one-way quantum information tasks, such as quantum
cryptography. The ``one-way EPR steering" or ``asymmetric EPR steering" is an
important ``open question" first proposed by Wiseman \emph{et al.}
in 2007~\cite{WJD07}. Very recently, the question has been answered
by Bowles \emph{et al.} \cite{Bowles}, who presented a simple class
of one-way steerable states in a two-qubit system with at least 13 projective
measurements (a linear 14-setting steering inequality was given explicitly in the work). The inspiring result for the first time theoretically confirms quantum
nonlocality can be fundamentally asymmetric. Later on,  Bowles \emph{et al.} investigated the one-way steering problem by
presenting a sufficient criterion (being a nonlinear criterion) for guaranteeing that a two-qubit state is unsteerable \cite{Bowles2} .

In this work, we focus on another curious quantum phenomenon raised by steering: Bell nonlocal states can be constructed from some
EPR steerable states. Explicitly we present a theorem, showing that for any two-qubit state $\tau$, if its corresponding state $\rho$ is EPR steerable, then the state $\tau$ must be Bell nonlocal. Bell's nonlocality of the quantum state $\tau$ can be detected indirectly by the violation of steering inequality for the quantum state $\rho$. The novel result not only pinpoints a deep connection between EPR steering and Bell's nonlocality, but also sheds a new light to avoid locality loophole in Bell's tests and make Bell's nonlocality easier for demonstration.
In addition, we also present a 9-setting linear steering inequality for developing more efficient one-way steering and detecting some Bell nonlocal states. We find that the new steering inequality can actually improve the result of \cite{Bowles} by detecting the one-way steering with fewer measurement settings but with larger quantum violations, which would be helpful for the experimenters.

%The second purpose of this work is to investigate Bell's nonlocality
%through EPR steering. We observe a new and curious quantum
%phenomenon called as the steering-induced Bell's nonlocality.
%Namely, we find that Bell nonlocal states can be induced from some
%EPR steerable states. This novel finding not only offers a
%distinctive way to study Bell's nonlocality without Bell's
%inequality but with steering inequality. Our results lead to a deeper connection between the steering and the
%Bell-nonlocality, and may have potential applications in the one-way
%quantum information processing.
%
%However, based on the obtained result, it is still difficult to observe such a kind of
%one-way steering phenomenon from the viewpoint of experiment, due to
%unrealized measurement error control and the tiny inequality
%violation. Consequently, one purpose of this work is to advance the
%study of unidirectional quantum steering. To do this, we
%propose a more efficient one-way steering. A
%9-setting linear steering inequality has also been presented
%explicitly for the experimental test. Based on this one may detect
%the one-way steering with fewer measurement settings but with larger
%quantum violations.

%\emph{More Efficient One-Way EPR Steering.}---

%\section{Experimental Proposal for More Efficient One-Way EPR Steering}

\vspace{2mm}

\noindent{\large{\bf Results}}

\vspace{2mm}

\noindent\textbf{Bell's Nonlocal states can be constructed from EPR steerable states.} It is well-known that quantum nonlocality possesses an interesting hierarchical structure (see Fig. 1). EPR steering is a weaker nonlocality in comparison to Bell's nonlocality.  Here we would like to pinpoint a curious quantum phenomenon directly connecting these two different types of nonlocality. We find that Bell's nonlocal states can be constructed from some EPR steerable states, which indicates that Bell's nonlocality can be detected indirectly
through EPR steering (see Fig. 2), and offers a distinctive way to study Bell's nonlocality. The result can be expressed as the following theorem.

%Naturally, if a state $\rho_{AB}$  violates a steering inequality
%(such as the inequality (\ref{inequality})), then it implies that
%$\rho_{AB}$ possesses the EPR steerability. However, the important
%application of steering inequalities is not restricted to detecting
%steerability itself, it can even be used to detect Bell's
%nonlocality. Here we explore the phenomenon of steering-induced
%Bell's nonlocality, which offers a distinctive way to study Bell's
%nonlocality.

\textbf{Theorem 1}: For any two-qubit state $\tau_{AB}$ shared by
Alice and Bob, define another state
\begin{eqnarray}\label{rhomu}
\rho_{AB}=\mu \; \tau_{AB} +(1-\mu) \tau'_{AB},
\end{eqnarray}
with $\tau'_{AB}={\tau }_{A}\otimes \openone/2$, ${\tau }_{A}=\textrm{tr}_{B}[\tau_{AB} ]=\textrm{tr}_{B}[\rho_{AB} ]$ being the reduced density matrix at Alice's side,
and $\mu=\frac{1}{\sqrt{3}}$. If $\rho_{AB}$ is EPR steerable, then $\tau_{AB} $ is Bell nonlocal.

\textbf{Proof.} The implication of the theorem is that, the EPR steerability of the
state $\rho_{AB}$ determines Bell's nonlocality of the state
$\tau_{AB} $. Namely, the nonexistence of local hidden state (LHS) model for $\rho_{AB}$
implies the nonexistence of LHV model for $\tau_{AB}$. We shall prove the theorem by
proving its converse negative proposition: if the state $\tau_{AB} $
has a LHV model description, then the state $\rho_{AB}$ has a LHS
model description.

Suppose $\tau_{AB}$ has a LHV model description, then by
definition£¬ for any projective measurements $A$ for Alice and $B$
for Bob, one always has the following relation
\begin{eqnarray}\label{LHV}
P(a,b|A, B, \tau_{AB} )=\sum_{\xi }P(a|A,\xi )P(b|B,\xi )P_{\xi }.
\end{eqnarray}
Here $P(a,b|A,B,\tau_{AB} )$ is the joint probability, quantum
mechanically it is computed as $P(a,b|A,B,\tau_{AB}
)=\textrm{tr}[(\hat{\Pi} ^{\hat{n}_A}_{a}\otimes \hat{\Pi}
^{\hat{n}_B}_{b})\; \tau_{AB} ]$,
%\begin{eqnarray}\label{Pab}
%P(a,b|A,B,\tau )=\textrm{tr}[(\hat{\Pi} ^{\hat{n}_A}_{a}\otimes
%\hat{\Pi} ^{\hat{n}_B}_{b})\; \tau ],
%\end{eqnarray}
$\hat{\Pi} ^{\hat{n}_A}_{a}$ is the projective measurement along the ${\hat{n}_A}$-direction with measurement outcome $a$ for Alice,
$\hat{\Pi} ^{\hat{n}_B}_{b}$ is the projective measurement along the ${\hat{n}_B}$-direction with measurement outcome $b$ for Bob (with $a, b=0, 1$),  $P(a|A,\xi
), P(b|B,\xi )$ and $P_{\xi }$  denote some (positive, normalized)
probability distributions.
% satisfying $\sum_{a=0}^{1}P(a|A,\xi )=1,
%\;\; \sum_{b=0}^{1}P(b|B,\xi )=1, \;\; \sum_{\xi }P_{\xi }=1.$
%\begin{eqnarray}
%\sum_{a=0}^{1}P(a|A,\xi )=1, \;\; \sum_{b=0}^{1}P(b|B,\xi )=1, \;\;
%\sum_{\xi }P_{\xi }=1.
%\end{eqnarray}

Let the measurement settings at Bob's side be picked out as ${x, y,
z}$. In this situation, Bob's projectors are $\hat{\Pi} ^{x}_{b}$,
$\hat{\Pi} ^{y}_{b}$, $\hat{\Pi} ^{z}_{b}$, respectively. Since the
state $\tau_{AB}$ has a LHV model description, based on Eq.
(\ref{LHV}) we explicitly have (with $\hat{n}= x, y, z$)
\begin{eqnarray}\label{ELHV}
P(a,0|A, \hat{n}, \tau_{AB} )&=&\sum_{\xi }^{}P(a|A,\xi )P(0|\hat{n},\xi )P^{}_{\xi },\nonumber\\
P(a,1|A, \hat{n}, \tau_{AB} )&=&\sum_{\xi }^{}P(a|A,\xi
)P(1|\hat{n},\xi )P^{}_{\xi }.
\end{eqnarray}

%\begin{subequations}  \label{ELHV}
%\begin{eqnarray}\label{ELHV}
%P(a,0|A, x, \tau_{AB} )&=&\sum_{\xi }^{}P(a|A,\xi )P(0|x,\xi )P^{}_{\xi },\nonumber\\
%P(a,1|A, x, \tau_{AB} )&=&\sum_{\xi }^{}P(a|A,\xi )P(1|x,\xi )P^{}_{\xi },\nonumber\\
%P(a,0|A, y, \tau_{AB} )&=&\sum_{\xi }^{}P(a|A,\xi )P(0|y,\xi )P^{}_{\xi },\nonumber\\
%P(a,1|A, y, \tau_{AB} )&=&\sum_{\xi }^{}P(a|A,\xi )P(1|y,\xi )P^{}_{\xi },\nonumber\\
%P(a,0|A, z, \tau_{AB} )&=&\sum_{\xi }^{}P(a|A,\xi )P(0|z,\xi )P^{}_{\xi },\nonumber\\
%P(a,1|A, z, \tau_{AB} )&=&\sum_{\xi }^{}P(a|A,\xi )P(1|z,\xi
%)P^{}_{\xi }.
%\end{eqnarray}
%\end{subequations}

We now turn to study the EPR steerability of $\rho_{AB}$. After
Alice performs the projective measurement on her qubit, the state
$\rho_{AB}$ collapses to Bob's conditional states (unnormalized) as
\begin{eqnarray}\label{pp}
\tilde{\rho}^{\hat{n}_A}_a={\rm tr}_A[(\hat{\Pi} ^{\hat{n}_A}_{a}
\otimes \openone) \rho_{AB}], \;\;\; a=0, 1.
\end{eqnarray}
To prove that there exists a LHS model for $\rho_{AB}$ is equivalent
to proving that, for any measurement $\hat{\Pi} ^{\hat{n}_A}_{a}$
and outcome $a$, one can always find a hidden state ensemble $\{
\wp_{\xi} \rho_{\xi} \}$ and the conditional probabilities
$\wp(a|\hat{n},\xi)$, such that the relation
\begin{eqnarray}\label{LHS1}
&&\tilde{\rho}^{\hat{n}_A}_a=\sum_{\xi} \wp(a|\hat{n}_A,\xi)
\wp_{\xi} \rho_{\xi},
\end{eqnarray}
is always satisfied. Here $\xi$'s are the local hidden variables,
$\rho_{\xi}$'s are the hidden states, $\wp_{\xi}$ and
$\wp(a|\hat{n},\xi)$ are probabilities satisfying
$\sum_\xi\wp_{\xi}=1$ and $\sum_a \wp(a|\hat{n}_A,\xi) =1$. If there
exist some specific measurement settings of Alice, such that Eq.
(\ref{LHS1}) cannot be satisfied, then one must conclude that the
state $\rho_{AB}$ is steerable (in the sense of Alice steers Bob's
particle).

Suppose there is a LHS model description for ${\rho }_{AB}$, then it
implies that, for Eq. (\ref{LHS1}) one can always find the solutions
of $\{\wp(a|\hat{n}_A,\xi), \wp_{\xi}, \rho_\xi\}$ if Eq.
(\ref{ELHV}) is valid. The solutions are given as follows:
%\begin{subequations}\label{LHS2}
\begin{eqnarray}\label{LHS2}
&&\wp(a|\hat{n}_A,\xi)=P(a|A,\xi ), \;\; \wp_{\xi}={P}_{\xi },\nonumber\\
&& \rho_\xi=\frac{\openone+\vec{\sigma}\cdot \vec{{r}}_{\xi }}{2},
\end{eqnarray}
%\end{subequations}
where $\openone$ is the $2\times 2$ identity matrix, $\vec\sigma=(\sigma_x,  \sigma_y, \sigma_z)$ is
the vector of the Pauli matrices, and the hidden state $\rho_\xi$ has been parameterized in the
Bloch-vector form, with
\begin{eqnarray}\label{r}
\vec{r}_{\xi }=\mu\; \left({2P(0|x,\xi )-1},\;{2P(0|y,\xi
)-1},\;{2P(0|z,\xi )-1}\right),
\end{eqnarray}
which is the Bloch vector for density matrix of a qubit. It can be
checked that $|\vec{r}_{\xi }|\leq 1$, and this ensures $\rho_\xi$ being
a density matrix.

%Equation (\ref{ELHV}) is helpful for proving the existence of a LHS model description, and the detail verification is given in Supplementary Materials.

%This ends the proof. \hfill \endproof

By substituting Eq. (\ref{LHS2}) into Eq. (\ref{LHS1}), we obtain
\begin{eqnarray}\label{LHS3}
\tilde{\rho}^{\hat{n}_A}_a=\sum_{\xi }^{}P(a|A,\xi ) {P}_{\xi }\;
\frac{\openone+\vec{\sigma}\cdot \vec{{r}}_{\xi }}{2}.
\end{eqnarray}
To prove the theorem is to verify the relation (\ref{LHS3}) is always satisfied
if Eq. (\ref{ELHV}) is valid. The verification can be found in \textbf{Methods}.

\emph{Remark 1.---} In Eq. (\ref{r}), by requiring the condition $|\vec{r}_{\xi }|\leq 1$ be valid for any probabilities $P(0|x,\xi ), P(0|y,\xi ), P(0|z,\xi ) \in [0,1]$, in general one can have $\mu\in[0,1/\sqrt{3}]$. Generally, Theorem 1 is valid for any $\mu\in[0,1/\sqrt{3}]$. In the theorem we have chosen the parameter $\mu$ as its maximal value $1/\sqrt{3}$, because the state $\tau_{AB}$ is convexed with a separable state $\tau'_{AB}$, the larger value of $\mu$, the easier to detect the EPR steerability.

In the following, we provide two examples for the theorem, showing that Bell's nonlocality of quantum states can be detected indirectly by the violations of some steering inequalities.

\emph{Example 1.---} For example, let us detect Bell's nonlocality
of the maximally entangled state (with ${\tau }_{AB}=|\Psi\rangle\langle \Psi|$)
%$|\Psi\rangle=\frac{1}{\sqrt{2}}(|00\rangle+ |11\rangle)$
\begin{eqnarray}\label{2qubitstate1}
&&|\Psi\rangle=\frac{1}{\sqrt{2}}(|00\rangle+ |11\rangle)
\end{eqnarray}
without Bell's inequality. Based on the theorem, it is equivalent to
detect the EPR steerability of the following two-qubit state
\begin{eqnarray}\label{2qubitstate2}
\rho_{AB}= \frac{1}{\sqrt{3}}\; |\Psi\rangle\langle \Psi| +(1-\frac{1}{\sqrt{3}})\; {\tau }_{A}\otimes \frac{\openone}{2},
\end{eqnarray}
with ${\tau }_{A}=\openone/2$. The state (\ref{2qubitstate2}) is
nothing but the Werner state \cite{werner89} with the visibility
equals to $1/\sqrt{3}$, its steerability can be tested  by using the
steering inequality proposed in Ref. \cite{NP2010} as
\begin{eqnarray}\label{sn}
\mathcal{S}_{N}=\frac{1}{N}\sum_{k=1}^N\langle
A_k\vec{\sigma}_k^B\rangle\leq C_N
\end{eqnarray}
with $N=6$. Here $\mathcal{S}_{N}$ is the steering parameter for
$N$ measurement settings, and $C_N$ is the classical bound, with
$C_6=(1 + \sqrt{5})/6\simeq 0.5393$. The maximal quantum violation
of the steering inequality is $\mathcal{S}_{6}^{\rm
max}=1/\sqrt{3}\simeq 0.5774$, which beats the classical bound.

\emph{Remark 2.---} In a two-qubit system, Bell's nonlocality is usually
detected by quantum violation of the Clause-Horne-Shimony-Holt
inequality~\cite{CHSH}. Bell's nonlocality
is the strongest type of nonlocality, due to this reason Bell-test
experiments have encountered both the locality loophole and the
detection loophole for a very long time~\cite{Hensen}. As a weaker
nonlocality, EPR steering naturally escapes from the locality
loophole and is correspondingly easier to be demonstrated without
the detection loophole~\cite{PRX}\cite{NJP}, as stated in
\cite{NP2010}: ``\emph{because the degree of correlation required
for EPR steering is smaller than that for violation of a Bell
inequality, it should be correspondingly easier to demonstrate
steering of qubits without making the fair-sampling assumption}
[i.e., closing the detection loophole]". Indeed, the steerability of
the Werner state has been experimentally detected in \cite{NP2010}
by the steering inequality (\ref{sn}). Our result shows that the EPR
steerability of the state $\rho_{AB}$ determines Bell's nonlocality
of the state $\tau_{AB} $, thus may shed a new light to realize a
loophole-free Bell-test experiment through the violation of steering
inequality.

\emph{Example 2.---} The theorem naturally provides a steering-based
criterion for Bell's nonlocality , which is expressed as follows: given an EPR steerable two-qubit state $\rho_{AB}$, if the matrix
\begin{eqnarray}\label{rhomup}
\tau_{AB}=\sqrt{3} \; \rho_{AB} -(\sqrt{3}-1) \tau'_{AB},
\end{eqnarray}
is a two-qubit density matrix, then $\tau_{AB} $ is Bell nonlocal.

Let us consider a two-qubit state $\rho_{AB}$ in the following form
\begin{eqnarray}\label{state-2}
\rho_{AB}= \frac{1}{4} \biggr(\openone\otimes \openone +\beta \sigma_{3} \otimes \openone +\gamma \openone\otimes \sigma_3
-\alpha\sum_{k=1}^3  \sigma_k
\otimes \sigma_k \biggr).
\end{eqnarray}
By substituting the state $\rho_{AB}$ as in Eq. (\ref{state-2}) into Eq. (\ref{rhomup}), then one obtains
\begin{eqnarray}\label{rhomup-2}
&&\tau_{AB}=\frac{1}{4} \biggr(\openone\otimes \openone +\beta' \sigma_{3} \otimes \openone +\gamma' \openone\otimes \sigma_3
-\alpha' \; \sum_{k=1}^3  \sigma_k
\otimes \sigma_k \biggr),
\end{eqnarray}
with
\begin{eqnarray}
&&\beta'=\beta, \;\; \gamma'=\sqrt{3}\; \gamma,\;\; \alpha'= \sqrt{3}\; \alpha.
\end{eqnarray}
It is worth to mention that the steering inequality (\ref{sn}) is applicable to show Bell's nonlocality of $\tau_{AB}$ for some parameters $\alpha', \beta', \gamma'$. Here we would like to show that the similar task can be done by other new steering inequalities. In the following, we present a 9-setting linear steering inequality as
\begin{align}\label{inequality}
\sum^{9}_{i=1}\sum^{3}_{j=1}s_{ij}\langle ab\rangle_{ij}+\sum^{9}_{i=1}s^{A}_{i}\langle a\rangle_{i}+\sum^{3}_{j=1}s^{B}_{j}\langle b\rangle_{j}\leq L,
\end{align}
here for convenient we have used the same notations as in \cite{Bowles} (where $(\sigma_1, \sigma_2, \sigma_3)$ is equivalent to $(\sigma_x, \sigma_y, \sigma_z)$ ). The inequality are characterized by matrices $\{\bf{S}, \bf{S^{A}}, \bf{S^{B}}\}$ with real coefficients
$s_{ij}$, $s^{A}_{i}$, and $s^{B}_{j}$, and the local bound is $L=1$ (see \textbf{Supplementary Materials}). The steering inequality (\ref{inequality}) may have other particular application for improving the result Ref. \cite{Bowles} by developing more efficient one-way steering, which we shall address in the coming section. But now we use it to detect Bell's nonlocality.

For example, let $\alpha'=0.96$, $\beta'=-1/5$, $\gamma'=1/6$, ones finds that $\tau_{AB}$ is a two-qubit state, and the steering inequality (\ref{inequality}) is violated by the state $\rho_{AB}$ (with the violation value $1.0064$), hence the Bell's nonlocality of state $\tau_{AB}$ can be revealed in this way indirectly by the steerability of the state $\rho_{AB}$.

\vspace{3mm}

\noindent\textbf{More efficient one-way
EPR steering.}  Under local unitary transformation (LUT), any
two-qubit state can be written in the following form \cite{Luo}
%\begin{eqnarray}\label{state-1}
%\rho_{AB}&=& \frac{1}{4} \biggr(\openone\otimes \openone + \beta\; \vec\sigma\cdot \hat{u} \otimes \openone  +
% \gamma\; \openone\otimes \vec\sigma\cdot \hat{v}\nonumber\\
%&& + \sum_{k}^3  t_{k} \sigma_k
%\otimes \sigma_k \biggr),
%\end{eqnarray}
\begin{eqnarray}\label{state-1}
\rho_{AB}= \frac{1}{4} \biggr(\openone\otimes \openone + \beta\; \vec\sigma\cdot \hat{u} \otimes \openone  +
 \gamma\; \openone\otimes \vec\sigma\cdot \hat{v}
 + \sum_{k=1}^3  t_{k} \sigma_k
\otimes \sigma_k \biggr),
\end{eqnarray}
with $\beta, \gamma, t_{k}$ being the real coefficients,   and $\hat{u}, \hat{v}$ the unit vectors.
Obviously, under LUT, the state $\rho_{AB}$ is said to be symmetric if and only if $\beta=\gamma$ and $\hat{u}=\hat{v}$. Let one consider a simple situation with
$t_1=t_2=t_3=-\alpha$, and $\hat{u}=\hat{v}=(0,0,1)$, then he obtains the two-qubit state $\rho_{AB}$ as in Eq. (\ref{state-2}). In such a case, if $\rho_{AB}$ is a one-way steerable state, then one must
have $\beta\neq \gamma$.

In Ref. \cite{Bowles}, the authors have chosen
$\beta=\frac{2(1-\alpha)}{5}$, $\gamma=-\frac{3(1-\alpha)}{5}$ and
used the SDP program to numerically prove that
the state $\rho_{AB}$ is a one-way steerable state (with at least 13
projective measurements): for $\alpha\leq 1/2$, the state $\rho_{AB}$ is unsteerable from Bob to
Alice, while for $\alpha \gtrsim 0.4983$ the state is steerable from
Alice to Bob when Alice performs $14$ projective measurements. An explicit
14-setting steering inequality has been also proposed to  conform
the one-way steerability, although for $\alpha=1/2$, the quantum
violation is tiny (only $1.0004$). The inspiring result for the
first time confirms that the nonlocality can be
fundamentally asymmetric. However, the tiny inequality violation as well as the 14 measurement settings give rise to the difficulty in experimental detection. To advance the study of unidirectional quantum steering, here we present a more efficient class of one-way steerable states by choosing
\begin{eqnarray}\label{state-3}
\beta=\frac{4\alpha(1-\alpha)}{3}, \;\;\gamma=-2\alpha(1-\alpha),
\end{eqnarray}
with $\alpha\in[0,1]$. The state $\rho_{AB}(\alpha)$ is entangled for $\alpha>0.3279$ . With the help of the SDP program, we found that in
the range $0.4846\lesssim\alpha\leq 1/2$, the state $\rho(\alpha)$ is one-way steerable within 10-setting measurements, thus this is more efficient than the previous result
in Ref. \cite{Bowles} ( For the detail derivation of more efficient one-way EPR steering see \textbf{Supplementary Materials}). Furthermore, we can extract an explicit 9-setting steering inequalities (\ref{inequality}) based on the SDP program. It can be verified directly that, for the state $\rho_{AB}(1/2)$, the quantum violation of 9-setting inequality (\ref{inequality}) is $\frac{119}{116}\backsimeq 1.0258>1$, hence demonstrating steering from Alice to Bob. Compared to the previous result \cite{Bowles}, the amount of violation is much larger but achieved with fewer measurements. To our knowledge, we do not know whether the quantum violation by inequality (\ref{inequality}) could be observed with current quantum technology. However, we believe that this result would be interesting and helpful for both theoretical and experimental physicists.

\vspace{2mm}

\noindent{\large{\bf Discussion}}

\vspace{2mm}

In this work, we have presented a theorem showing that Bell nonlocal states can be constructed from some EPR steerable states. This result
not only offers a novel and distinctive way to study Bell's nonlocality with the violation of steering inequality, but also may avoid locality loophole in Bell's tests and make Bell's nonlocality easier for demonstration. An interesting and inverse problem is whether one can construct some steerable states $\tau_{AB}$ from some Bell nonlocal state $\rho_{AB}$, because Bell's nonlocality has been researched more deeply in theoretical aspect, so that people can conveniently study steering via known criteria of Bell's nonlocality. Furthermore, an explicit 9-setting linear steering inequality has also been presented for detecting some Bell nonlocal states and developing more efficient one-way steering. This result allows one to observe one-way EPR steering with fewer measurement setting but with larger quantum violations. We hope experimental progress in this direction could be made in the near future.

\vspace{2mm}

\noindent{\large{\bf Methods}}

\vspace{2mm}

\noindent\textbf{Verification of equation (\ref{LHS3}).}
%\noindent\emph{Calculation of the left-hand side of Eq. (\ref{LHS3}).}---
Let us calculate the left-hand side of Eq. (\ref{LHS3}). One
has

%\begin{subequations}
\begin{eqnarray*}\label{left1}
&&\tilde{\rho}^{\hat{n}_A}_a
={\rm tr}_A[(\hat{\Pi} ^{\hat{n}_A}_{a} \otimes \openone) \rho_{AB}] \nonumber\\
&&={\rm tr}_A[(\hat{\Pi} ^{\hat{n}_A}_{a} \otimes \openone) (\mu \; \tau_{AB} +(1-\mu) \tau'_{AB})] \\
%&&=\mu\;{\rm tr}_A[(\hat{\Pi} ^{\hat{n}_A}_{a} \otimes \openone)  \tau_{AB} ]+(1-\mu)\;{\rm tr}_A[(\hat{\Pi} ^{\hat{n}_A}_{a} \otimes \openone) ({\tau }_{A}\otimes \frac{\openone}{2})]\nonumber\\
%&&=\mu\;{\rm tr}_A[(\hat{\Pi} ^{\hat{n}_A}_{a} \otimes \openone)  \tau_{AB} ]+(1-\mu)\;{\rm tr}[\hat{\Pi} ^{\hat{n}_A}_{a} {\tau }_{A} ]\;\frac{\openone}{2}\nonumber\\
&&=\mu\;{\rm tr}_A[(\hat{\Pi} ^{\hat{n}_A}_{a} \otimes \openone)
\tau_{AB} ]+(1-\mu)\;P(a|A, \tau_{AB}
)\;\frac{\openone}{2},\nonumber
\end{eqnarray*}
%\end{subequations}
where $P(a|A, \tau_{AB})={\rm tr}[\hat{\Pi} ^{\hat{n}_A}_{a} {\tau
}_{A} ]$ is the marginal probability of Alice when she measures $A$
and gets the outcome $a$. For convenient, let us denote the $2\times
2$ matrix $\tilde{\rho}^{\hat{n}_A}_a$ as
\begin{eqnarray*}\label{left}
\tilde{\rho}^{\hat{n}_A}_a=\begin{bmatrix}
 {\nu}_{11}&{\nu}_{12} \\
 {\nu}_{21}&{\nu}_{22}
\end{bmatrix},
\end{eqnarray*}
and calculate its each element. We get
%\begin{subequations}
\begin{eqnarray*}\label{element11}
&&{\nu}_{11}=\textrm{tr}\left[\begin{bmatrix}
 1&0 \\
 0&0
\end{bmatrix}\begin{bmatrix}
{\nu}_{11}&{\nu}_{12} \\
 {\nu}_{21}&{\nu}_{22}
\end{bmatrix}\right]=\textrm{tr}[\hat{\Pi}^{z}_{0}\; \tilde{\rho}^{\hat{n}_A}_a]\nonumber\\
%&&=\textrm{tr}[\hat{\Pi}^{z}_{0}\; \{\mu\;{\rm tr}_A[(\hat{\Pi} ^{\hat{n}_A}_{a} \otimes \openone)  \tau_{AB} ]+(1-\mu)\;P(a|A, \tau_{AB} )\;\frac{\openone}{2}\}]\nonumber\\
%&&=\mu\; \textrm{tr}[(\hat{\Pi} ^{\hat{n}_A}_{a} \otimes
%\hat{\Pi}^{z}_{0})  \tau_{AB} ]+
%(1-\mu)\;P(a|A, \tau_{AB} )\;\textrm{tr}[\hat{\Pi}^{z}_{0}\; (\frac{\openone}{2})]\nonumber\\
&&=\mu\; P(a,0|A,z,\tau_{AB} )+(1-\mu) P(a|A, \tau_{AB}
)\frac{1}{2},
\end{eqnarray*}
%\end{subequations}
and similarly,
\begin{eqnarray*}\label{element22}
&&{\nu}_{22}=\textrm{tr}\left[\begin{bmatrix}
 0&0 \\
 0&1
\end{bmatrix}\begin{bmatrix}
{\nu}_{11}&{\nu}_{12} \\
 {\nu}_{21}&{\nu}_{22}
\end{bmatrix}\right]=\textrm{tr}[\hat{\Pi}^{z}_{1}\; \tilde{\rho}^{\hat{n}_A}_a]
=\mu\; P(a,1|A,z,\tau_{AB} )+(1-\mu) P(a|A, \tau_{AB}
)\frac{1}{2}.
\end{eqnarray*}

Note that
${\nu}_{11}+{\nu}_{22}=\textrm{tr}[\tilde{\rho}^{\hat{n}_A}_a]=P(a|A,
\tau_{AB}),$
%\begin{eqnarray*}
%{\nu}_{11}+{\nu}_{22}=\textrm{tr}[\tilde{\rho}^{\hat{n}_A}_a]=P(a|A,
%\tau_{AB}),
%\end{eqnarray*}
we then have
\begin{eqnarray*}
{\nu}_{22}=-\mu\; P(a,0|A,z,\tau_{AB} )+(1+\mu) P(a|A, \tau_{AB}
)\frac{1}{2}.
\end{eqnarray*}
Because
\begin{eqnarray*}
\textrm{tr}\left[\begin{bmatrix}
 \frac{1}{2}&\frac{1}{2} \\
 \frac{1}{2}&\frac{1}{2}
\end{bmatrix}\tilde{\rho}^{\hat{n}_A}_a \right]
=\frac{1}{2}P(a|A,\tau_{AB} )+\textrm{Re}[{\nu}_{12}],
\end{eqnarray*}
with $\textrm{Re}[{\nu}_{12}]$ is the real part of ${\nu}_{12}$,
thus,
\begin{eqnarray*}
&&\textrm{Re}[{\nu}_{12}]=\textrm{tr}[\hat{\Pi}^{x}_{0} \tilde{\rho}^{\hat{n}_A}_a]-\frac{1}{2}P(a|A,\tau_{AB})
%&&=\mu\; P(a,1|A,x,\tau_{AB} )+(1-\mu) P(a|A, \tau_{AB} )\frac{1}{2}-\frac{1}{2}P(a|A,\tau_{AB})\nonumber\\
=\mu\; P(a,1|A,x,\tau_{AB} )-\frac{\mu}{2} P(a|A, \tau_{AB} ).
\end{eqnarray*}

Similarly, because
\begin{eqnarray*}
\textrm{tr}\left[\begin{bmatrix}
 \frac{1}{2}&-\frac{i}{2} \\
 \frac{i}{2}&\frac{1}{2}
\end{bmatrix}\tilde{\rho}^{\hat{n}_A}_a \right]
%=\frac{1}{2}({\nu}_{11}+{\nu}_{22})+\frac{i}{2}({\nu}_{12}-{\nu}_{21})
=\frac{1}{2}P(a|A,\rho )-\textrm{Im}[{\nu}_{12}],
\end{eqnarray*}
with $\textrm{Im}[{\nu}_{12}]$ is the imaginary part of
${\nu}_{12}$, thus,
\begin{eqnarray*}
&&\textrm{Im}[{\nu}_{12}]=-\textrm{tr}[\hat{\Pi}^{y}_{0} \tilde{\rho}^{\hat{n}_A}_a]+\frac{1}{2}P(a|A,\tau_{AB})
%&&=-\mu\; P(a,1|A,y,\tau_{AB} )-(1-\mu) P(a|A, \tau_{AB} )\frac{1}{2}+\frac{1}{2}P(a|A,\tau_{AB})\nonumber\\
=-\mu\; P(a,1|A,y,\tau_{AB} )+\frac{\mu}{2} P(a|A, \tau_{AB} ).
\end{eqnarray*}

By combining the above equations, we finally have

\begin{eqnarray}\label{leftnew}
&&\tilde{\rho}^{\hat{n}_A}_a=\begin{bmatrix}
 {\nu}_{11}&{\nu}_{12} \\
 {\nu}_{21}&{\nu}_{22}
\end{bmatrix}=\frac{{\nu}_{11}+{\nu}_{22}}{2}\;\openone+ \textrm{Re}[{\nu}_{12}]
\;\sigma_x -\textrm{Im}[{\nu}_{12}]
\;\sigma_y+\frac{{\nu}_{11}+{\nu}_{22}}{2}\;\sigma_z.
\end{eqnarray}

%\emph{Calculation of the right-hand side of Eq. (\ref{LHS3}).}---
Let us calculate the right-hand side of Eq. (\ref{LHS3}).
It gives
{\small \begin{eqnarray*} &&\sum_{\xi }^{}P(a|A,\xi ) {P}_{\xi }\;
\frac{\openone+\vec{\sigma}\cdot \vec{{r}}_{\xi }}{2}
%&&=\frac{\openone}{2}\; (\sum_{\xi }^{}P(a|A,\xi ) {P}_{\xi })+\frac{\mu}{2}\;\sum_{\xi }^{}P(a|A,\xi ) {P}_{\xi } \times\nonumber\\
%&& [\sigma_{x}(2P(0|x, \xi)-1)+\sigma_{y}(2P(0|y, \xi)-1)+\sigma_{z}(2P(0|z, \xi)-1)]\nonumber\\
=(\sum_{\xi }^{}P(a|A,\xi ) {P}_{\xi })\;\frac{\openone}{2} \nonumber\\
&&+\mu\; (\sum_{\xi }^{}P(a|A,\xi )P(0|x,\xi ){P}_{\xi}){\sigma}_{x}
-\frac{\mu}{2} (\sum_{\xi }^{}P(a|A,\xi ) {P}_{\xi }) \; \sigma_{x}\nonumber\\
&&+\mu\; (\sum_{\xi }^{}P(a|A,\xi )P(0|y,\xi ){P}_{\xi}){\sigma}_{y}-\frac{\mu}{2} (\sum_{\xi }^{}P(a|A,\xi ) {P}_{\xi }) \; \sigma_{y}\nonumber\\
&&+\mu\; (\sum_{\xi }^{}P(a|A,\xi )P(0|z,\xi
){P}_{\xi}){\sigma}_{z}-\frac{\mu}{2} (\sum_{\xi }^{}P(a|A,\xi )
{P}_{\xi }) \; \sigma_{z}.
\end{eqnarray*}
}
 With the help of Eq. (\ref{ELHV}) and using $\sum_{\xi }^{}P(a|A,\xi ) {P}_{\xi }=P(a|A,\tau_{AB}),$
%\begin{eqnarray*}
%\sum_{\xi }^{}P(a|A,\xi ) {P}_{\xi }=P(a|A,\tau_{AB}),
%\end{eqnarray*}
we finally have
\begin{eqnarray}\label{right}
&&\sum_{\xi }^{}P(a|A,\xi ) {P}_{\xi }\;
\frac{\openone+\vec{\sigma}\cdot \vec{{r}}_{\xi }}{2}
=P(a|A,\tau_{AB})\frac{\openone}{2}\nonumber\\
&&+\mu\; P(a,0|A,x,\tau_{AB}){\sigma}_{x}-\frac{\mu}{2}P(a|A,\tau_{AB}){\sigma}_{x}\nonumber\\
&&+\mu\;
P(a,0|A,y,\tau_{AB}){\sigma}_{y}-\frac{\mu}{2}P(a|A,\tau_{AB}){\sigma}_{y}\nonumber\\
&&+\mu\;
P(a,0|A,z,\tau_{AB}){\sigma}_{z}-\frac{\mu}{2}P(a|A,\tau_{AB}){\sigma}_{z}.
\end{eqnarray}
By comparing  Eq. (\ref{leftnew}) and Eq. (\ref{right}), it is easy
to see that Eq. (\ref{LHS3}) holds.  Thus, if there is a LHV model
description for $\tau_{AB}$, then there is a LHS model description
for $\rho_{AB}$. This completes the proof.

%\hfill \endproof

\vspace{2mm}

 \noindent{\bf Acknowledgements}

\noindent J.L.C. is supported by the National Basic Research Program
(973 Program) of China under Grant No.\ 2012CB921900 and the Natural
Science Foundations of China (Grant No.\ 11475089). C.R.
acknowledges supported by Youth Innovation Promotion Association
(CAS) No. 2015317, Natural Science Foundations of Chongqing
(No.cstc2013jcyjC00001, cstc2015jcyjA00021) and The
Project-sponsored by SRF for ROCS-SEM (No.Y51Z030W10). C.C. was
partially supported by NSFC (11301524, 11471307) and CSTC
(cstc2015jcyjys40001). A.K.P. is supported by the Special Project of
University of Ministry of Education of China and the Project of K.
P. Chair Professor of Zhejiang University of China.

\vspace{2mm}

 \noindent{\bf Author contributions}

\noindent J.L.C. initiated the idea. J.L.C, C.R., C.C. and X.J.Y.
derived the results. J.L.C. prepared the figure. J.L.C. and A.K.P.
wrote the main manuscript text. All authors reviewed the manuscript.

\vspace{2mm}

 \noindent{\bf Additional information}

\noindent\textbf{Competing financial interests:} The authors declare no
competing financial interests.\\

\newpage

\begin{figure}[t]
\includegraphics[width=80mm]{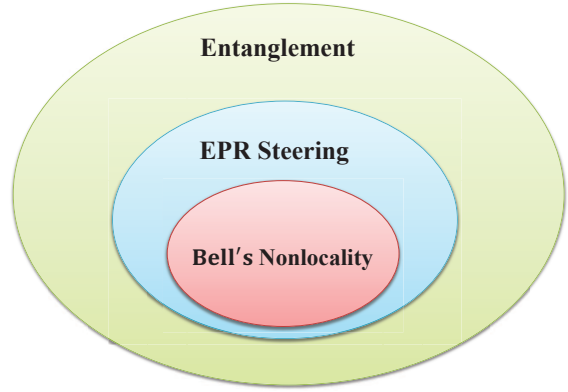}\\
\caption{\textbf{Hierarchical structure of quantum nonlocality.}  Bell's nonlocality is the strongest type of quantum nonlocality. If a state possesses EPR steerability or Bell's nonlocality, then the state must be entangled. EPR steering is a form of nonlocality intermediate between entanglement and Bell nonlocality.}\label{fig1}
\end{figure}

\newpage

\begin{figure}[t]
\includegraphics[width=80mm]{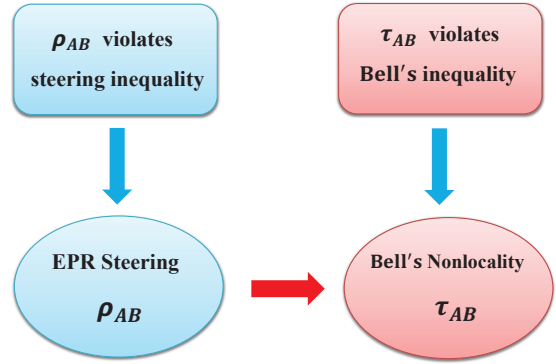}\\
\caption{\textbf{Illustration of detecting Bell's nonlocality
through EPR steering.} If a state $\rho_{AB}$  violates a steering
inequality, then it implies that $\rho_{AB}$ possesses the EPR
steerability. Traditionally, Bell's nonlocality of the two-qubit
state $\tau_{AB}$ is revealed by violations of Bell's inequality.
Based on Theorem 1, Bell's nonlocality of the state $\tau_{AB}$ can
be detected through EPR steerability of the state $\rho_{AB}$, and
the relation between $\rho_{AB}$ and $\tau_{AB}$ is given in Eq.
(\ref{rhomu}).}\label{fig2}
\end{figure}

\end{document}

\title{{\Large Supplementary Materials}  \\ \vspace{8pt} for ``Bell's Nonlocality Can be Detected by the Violation of Einstein-Podolsky-Rosen Steering Inequality"}

\author{Jing-Ling Chen}
 \affiliation{Theoretical Physics Division, Chern Institute of Mathematics, Nankai University,
 Tianjin 300071, People's Republic of China}
 \affiliation{Centre for Quantum Technologies, National University of Singapore,
 3 Science Drive 2, Singapore 117543, Singapore}

\author{Changliang Ren}
 \affiliation{Center for Nanofabrication and System Integration, Chongqing Institute of Green and Intelligent Technology, Chinese Academy of Sciences, Chongqing 400714, People's Republic of China}

\author{Changbo Chen}
 \affiliation{Chongqing Key Laboratory of Automated Reasoning and Cognition, Chongqing Institute of Green and Intelligent Technology, Chinese Academy of Sciences, Chongqing 400714, People's Republic of China}

\author{Xiang-Jun Ye}
 \affiliation{Key Laboratory of Quantum Information, University of Science and Technology of China, University of Science and Technology of China, Hefei 230026, People's Republic of China}
\affiliation{Synergetic Innovation Center of Quantum Information and Quantum Physics, University of Science and Technology of China, Hefei 230026, People's Republic of China}

\author{Arun Kumar Pati}
 \affiliation{Quantum Information and Computation Group, Harish-Chandra Research Institute, Chhatnag Road, Jhunsi, Allahabad 211019, India}

\date{\today}
%%%%%%%%%%%%%%%%%%%%%%%%%%%%%%%%%%%%%%%%%%%%%%%%%%%%%%%%%%%%%%%%%%%

%\pacs{03.65.Ud, 03.67.Mn, 42.50.Xa}

\maketitle

\section{I. Detail derivation of more efficient one-way EPR steering}

In Ref. \cite{Bowles}, Bowles, Vertesi, Quintino, and Brunner (BVQB) have presented a class of one-parameter two-qubit state
\begin{eqnarray}\label{brunnerstate}
\rho_{AB}(\alpha)=\alpha |\psi^-\rangle\langle\psi^-|+\frac{1-\alpha}{5}\biggr(2|0\rangle\langle0|\otimes\frac{\openone}{2}+3 \frac{\openone}{2}\otimes |1\rangle\langle1|\biggr),
\end{eqnarray}
where
\begin{eqnarray}\label{single}
|\psi^-\rangle=\frac{1}{2}(|01\rangle-|10\rangle)
\end{eqnarray}
is the singlet state and the parameter
\begin{eqnarray}\label{parameter}
\alpha\in [0,1].
\end{eqnarray}
%The state (\ref{brunnerstate}) has a definite meaning: it is a convex sum of the singlet state $|\psi^-\rangle\langle\psi^-|$ and a certain noise density matrix
%\begin{eqnarray}\label{noise1}
%\rho_{\rm noise}^{\rm BVQB}=\frac{1}{5}\biggr(2|0\rangle\langle0|\otimes\frac{\openone}{2}+3 \frac{\openone}{2}\otimes |1\rangle\langle1|\biggr).
%\end{eqnarray}
It is easy to find that the state (\ref{brunnerstate}) is identical to the following form of the two-qubit density matrix
\begin{eqnarray}\label{state-2}
\rho_{AB}= \frac{1}{4} \biggr(\openone\otimes \openone +\beta \sigma_{3} \otimes \openone +\gamma \openone\otimes \sigma_3
-\alpha\sum_{k=1}^3  \sigma_k
\otimes \sigma_k \biggr).
\end{eqnarray}
with the specific values
\begin{eqnarray}\label{ba1}
\beta=\frac{2(1-\alpha)}{5}, \;\;\; \gamma=-\frac{3(1-\alpha)}{5}.
\end{eqnarray}
However, BVQB did not mention where the state (\ref{brunnerstate}) came from and how to construct it. Here, we provide a detail derivation, and from the derivation one can naturally achieve some more efficient states for demonstrating one-way EPR steering.

The derivation is just based on the BVQB LHS model, in which Bob can never steer Alice with any measurement settings (see the section ``\emph{No steering from B to A}'' in \cite{Bowles}). In the BVQB model, supposed that Alice chooses an arbitrary measurement direction $\vec{x}=(x_1,x_2,x_3)$, and $\vec{y}=(y_1,y_2,y_3)$ for Bob, then from the viewpoint of LHS, the local expectation values and the correlation are given by
\begin{eqnarray}\label{lhs1}
\langle a\rangle_{\mathrm{LHS}}&=& \frac{x_3}{3},\nonumber\\
\langle b\rangle_{\mathrm{LHS}}&=&  -\frac{y_3}{2},\nonumber\\
\langle ab\rangle_{\mathrm{LHS}}&=& - \frac{\vec{x}\cdot\vec{y}}{2}.
\end{eqnarray}
Under local unitary transformation (LUT), any
two-qubit state can be written in the following form \cite{Luo}
\begin{eqnarray}\label{state-1}
\rho_{AB}= \frac{1}{4} \biggr(\openone\otimes \openone + \beta\; \vec\sigma\cdot \hat{u} \otimes \openone  +
 \gamma\; \openone\otimes \vec\sigma\cdot \hat{v}
 + \sum_{k=1}^3  t_{k} \sigma_k
\otimes \sigma_k \biggr),
\end{eqnarray}
with $\beta, \gamma, t_{k}$ being the real coefficients,   and $\hat{u}, \hat{v}$ the unit vectors.
During demonstrating the steerable states, we may ask a reverse question: for the BVQB model, which quantum states can be described by it? Without loss of generality, we can analyze how to extract these unsteerable states that can be described by the BVQB model from an arbitrary two-qubit state (\ref{state-1}). The necessary condition is that the quantum expectation values of arbitrary measurement directions should coincident with those of the BVQB model. Obviously, the joint and marginal expectation values derived from quantum mechanics should satisfy the following relations:
\begin{eqnarray}
\langle a\rangle_{\mathrm{QM}}&=&\tr[\rho_{AB}(\vec\sigma\cdot \vec{x}\otimes \openone)] \sim x_3,\nonumber\\
\langle b\rangle_{\mathrm{QM}}&=&\tr[\rho_{AB}(\openone\otimes\vec\sigma\cdot \vec{y})]\sim y_3,\nonumber\\
\langle ab\rangle_{\mathrm{QM}}&=&\tr[\rho_{AB}(\vec\sigma\cdot \vec{x}\otimes \vec\sigma\cdot \vec{y})]\sim\vec{x}\cdot\vec{y}.
\end{eqnarray}
This condition is satisfied if and only if $t_1=t_2=t_3=-\alpha$ and $\hat{u}=\hat{v}=(0,0,1)$, hence from the state (\ref{state-1}) we arrive at
\begin{eqnarray}\label{state-2b}
\rho_{AB}= \frac{1}{4} \biggr(\openone\otimes \openone +\beta \sigma_{3} \otimes \openone +\gamma \openone\otimes \sigma_3
-\alpha\sum_{k=1}^3  \sigma_k
\otimes \sigma_k \biggr)
\end{eqnarray}
which is just (\ref{state-2}).

For the state $\rho_{AB}(1/2)$ in (\ref{brunnerstate}) (It is sufficient to consider $\rho_{AB}(\alpha)$ with $\alpha=1/2$. The extension of the case $\alpha \le 1/2$ is straightforward \cite{Bowles}), one can have the joint and marginal quantum expectation values as
\begin{eqnarray}\label{qmbrunner}
\langle a\rangle_{\mathrm{QM}}&=&\tr[\rho_{AB}(\vec\sigma\cdot \vec{x}\otimes \openone)] = \frac{x_3}{5},\nonumber\\
\langle b\rangle_{\mathrm{QM}}&=&\tr[\rho_{AB}(\openone\otimes\vec\sigma\cdot \vec{y})] = -\frac{3 y_3}{10},\nonumber\\
\langle ab\rangle_{\mathrm{QM}}&=&\tr[\rho_{AB}(\vec\sigma\cdot \vec{x}\otimes \vec\sigma\cdot \vec{y})] = -\frac{\vec{x}\cdot\vec{y}}{2}.
\end{eqnarray}
However, Eq. (\ref{lhs1}) cannot simulate directly the quantum values expressed in Eq. (\ref{qmbrunner}). To do this, BVQB introduced a parameter of flipping probability $f$ with
\begin{eqnarray}\label{f12}
f\in[0,1/2],
\end{eqnarray}
then  Eq. (\ref{lhs1}) becomes
\begin{eqnarray}\label{lhs2}
\langle a\rangle_{\mathrm{LHS}}&=& \frac{1-2f}{3}x_3,\nonumber\\
\langle b\rangle_{\mathrm{LHS}}&=&  -\frac{1-2f}{2}y_3,\nonumber\\
\langle ab\rangle_{\mathrm{LHS}}&=& - \frac{\vec{x}\cdot\vec{y}}{2}.
\end{eqnarray}
By choosing $f=1/5$, Eq. (\ref{lhs1}) exactly simulates the quantum results in Eq. (\ref{qmbrunner}), thus proving $\rho_{AB}(1/2)$ is unsteerable from Bob to Alice.

Let us return to the state (\ref{state-2b}), quantum mechanically one can have
\begin{eqnarray}\label{qmnew}
\langle a\rangle_{\mathrm{QM}}&=&\tr[\rho_{AB}(\vec\sigma\cdot \vec{x}\otimes \openone)] = \beta\; x_3,\nonumber\\
\langle b\rangle_{\mathrm{QM}}&=&\tr[\rho_{AB}(\openone\otimes\vec\sigma\cdot \vec{y})] = \gamma\; y_3,\nonumber\\
\langle ab\rangle_{\mathrm{QM}}&=&\tr[\rho_{AB}(\vec\sigma\cdot \vec{x}\otimes \vec\sigma\cdot \vec{y})] = -\alpha\; \vec{x}\cdot\vec{y}.
\end{eqnarray}
By comparing the first two formulae in Eq. (\ref{lhs2}) and Eq. (\ref{qmnew}) one has
\begin{eqnarray}\label{betaga}
\beta=\frac{1-2f}{3}, \;\;\gamma=-\frac{1-2f}{2}.
\end{eqnarray}
Submitting Eq. (\ref{betaga}) into Eq. (\ref{state-2b}) one has a two-parameter quantum state as
\begin{eqnarray}\label{state-2c}
\rho_{AB}(\alpha, f) = \frac{1}{4} \biggr(\openone\otimes \openone +\frac{(1-2f)}{3} \;\sigma_{3} \otimes \openone -\frac{(1-2f)}{2}\; \openone\otimes \sigma_3
-\alpha\sum_{k=1}^3  \sigma_k
\otimes \sigma_k \biggr).
\end{eqnarray}

\emph{Remark 1.---} If one chooses
\begin{eqnarray}\label{betaga1}
f=1/2,
\end{eqnarray}
then one has $\beta=\gamma=0$. In this case, the state (\ref{state-2c}) is a symmetric state, and it is just the Werner state. It is well-known that the Werner state is unsteerable for the region $\alpha\leq 1/2$ \cite{WJD07}.

\emph{Remark 2.---} By comparing Eq. (\ref{ba1}) and Eq. (\ref{betaga}), one obtains
\begin{eqnarray}\label{f2}
f=\frac{3}{5}\biggr( \alpha-\frac{1}{6}\biggr).
\end{eqnarray}
Namely, if one selects the parameter $f$ as a linear function of $\alpha$ as $f(\alpha)=3(\alpha-1/6)/5$, then one recovers the BVQB state as in Eq. (\ref{brunnerstate}).
If $\alpha$ runs from $1/6$ to $1$, the parameter $f$ will run from $0$ to $1/2$, and $\alpha=1/2$ corresponds to $f=1/5$. For the state $\rho_{AB}(\alpha=1/2, f=1/5)$, Ref. \cite{Bowles} has proved that with at lease 13 projective measurements Alice can steer Bob's qubit state. An explicit 14-setting steering inequality has also been proposed in Ref. \cite{Bowles} to  conform the one-way steerability, although for $\alpha=1/2$, the quantum violation is tiny (only $2269/2268 \backsimeq1.0004$). The inspiring result for the first time confirms that the nonlocality can be fundamentally asymmetric. However, the tiny inequality violation as well as the 14 measurement settings give rise to the difficulty in experimental detection. By the way, the BVQB LHS model is not valid for the region of $\alpha\in[0,\frac{1}{6}]$, however in this case $\rho_{AB}(\alpha)$ is a separable state that can easily have other description of LHS models.

\emph{Remark 3.---} We now come to extract some more efficient one-way steerable states from  Eq.(\ref{state-2c}). For convenient to compare with the result of Ref. \cite{Bowles}, here we also present a one-parameter quantum state by choosing
\begin{eqnarray}\label{ourf1}
f=2\biggr(\alpha-\frac{1}{2}\biggr)^2.
\end{eqnarray}
Explicitly, the state is given by
\begin{eqnarray}\label{state-2d}
\rho_{AB}(\alpha) = \frac{1}{4} \biggr(\openone\otimes \openone +\frac{4\alpha(1-\alpha)}{3} \;\sigma_{3} \otimes \openone -2\alpha(1-\alpha)\; \openone\otimes \sigma_3
-\alpha\sum_{k=1}^3  \sigma_k
\otimes \sigma_k \biggr).
\end{eqnarray}
For the selection of the parameter $f$ in Eq. (\ref{ourf1}), there are three reasons: (i) similar to the state presented in Ref. \cite{Bowles}, the matrix $\rho_{AB}(\alpha)$ in Eq. (\ref{state-2d}) is always a density matrix for the all region of $\alpha\in[0,1]$; (ii) when $\alpha$ runs from $0$ to $1$, the parameter $f$ always stays in the region of $[0,1/2]$, such that if $\rho_{AB}(\alpha=1/2)$ has a description of the BVQB LHS model, then it ensures that $\rho_{AB}(\alpha<1/2)$ also has a description of the BVQB model; (iii) $\rho_{AB}(\alpha)$ is a more  efficient one-way steerable state, as we shall show below.

First, based on the BVQB model, no steering from Bob to Alice for the state $\rho_{AB}(\alpha)$ when $\alpha \leq 1/2$. Second, we need to show Alice can steer Bob with the state $\rho_{AB}(\alpha \leq 1/2)$.  We shall show that the state $\rho(\alpha)$ with $\alpha\gtrsim 0.4846$ is steerable from Alice to Bob. With the help of the SDP program, we have calculated the threshold
values $\alpha^{*}$ for which the state $\rho(\alpha)$ is steerable from Alice to Bob for different $m$ measurement directions $\vec{x}$ with $m=2,3,...,10$ (see Table I).
Definitely, for $m=8$ we obtain $\alpha^{*}\backsimeq 0.4982$, thus implying that the state $\rho(\alpha)$ with $\alpha^{*}\gtrsim 0.4982$ is steerable from Alice to Bob.
Hence it shows a class of  more efficient one-way steerable states than that of Ref. \cite{Bowles}, in which the states with $\alpha^{*}\gtrsim 0.4983$ are steerable from Alice to Bob
when Alice performs $14$ projective measurements. And for $m=10$, we obtain a larger value of $\alpha^{*}\backsimeq 0.4846$.
\begin{table}[h]
%\scriptsize
%\small
%\centering
  \begin{tabular}{ c c c c c c c c c c} \hline \hline
   $m$ & 2 & 3 & 4 & 5 & 6 & 7 & 8 & 9&10\\ \hline
   $\alpha^{*} $ &  0.6302& 0.5461 & 0.5244 & 0.5147 & 0.5071 & 0.5041 & 0.4982 & 0.4855& 0.4846\\
    \hline \hline
   \end{tabular}
  \caption{\label{Table1} Threshold values $\alpha^{*}$ for which the state $\rho(\alpha)$ is steerable from Alice to Bob, when Alice performs $m=2,3,...,10$ projective measurements on her qubit, respectively.  }
\end{table}

Third, we can extract analytic 8-setting and 9-setting linear steering inequalities based on SDP program. For example, the 8-setting steering inequality is given by
\begin{align}\label{8inequality}
\sum^{8}_{i=1}\sum^{3}_{j=1}s_{ij}\langle ab\rangle_{ij}+\sum^{8}_{i=1}s^{A}_{i}\langle a\rangle_{i}+\sum^{3}_{j=1}s^{B}_{j}\langle b\rangle_{j}\leq L,
\end{align}
with the local bound $L=1$, and
\begin{eqnarray}
&& \bf{S} = \left( \begin{array}{ccc}
-\frac{1}{75} & -\frac{6}{43} & -\frac{13}{86}\\
   -\frac{16}{87} & \frac{9}{157} & \frac{13}{111}\\
  -\frac{1}{314} & -\frac{15}{116} & -\frac{1}{35}\\
   -\frac{18}{103} & \frac{8}{67} & -\frac{1}{46}\\
   -\frac{2}{49} & \frac{15}{79} & -\frac{16}{109}\\
   \frac{15}{119} & \frac{15}{133} & -\frac{14}{79}\\
   \frac{5}{117} & -\frac{1}{104} & \frac{13}{60}\\
   -\frac{22}{103} & -\frac{27}{242} & -\frac{7}{66}\end{array} \right),
 \bf{S^{A}}= \left( \begin{array}{c}
 -\frac{14}{209} \\
   \frac{2}{39} \\
  -\frac{1}{79} \\
   -\frac{1}{98}\\
   -\frac{5}{77} \\
   -\frac{10}{139} \\
   \frac{7}{73}\\
   -\frac{1}{21}\end{array} \right),
 \bf{S^{B}}=\left( \begin{array}{c}
 -\frac{1}{71} \\
   \frac{1}{1888} \\
  -\frac{75}{173} \end{array} \right).
  \end{eqnarray}
It can be verified that the state $\rho_{AB}(\alpha=1/2)$ violates the 8-setting inequality with violation value as $\frac{313}{312}\backsimeq 1.0032$, where the 8 measurement settings of Alice can be characterized by Bloch vectors $\vec{x}_{i}$ ($i=1, 2,...,8$), which are
\begin{eqnarray}
\bf{V}=\begin{pmatrix}
\frac{5}{66} & -\frac{65}{82} & \vert z_1\vert \\
   \frac{57}{65} & -\frac{38}{139} & -\vert z_2\vert \\
  \frac{1}{41} & \frac{81}{82} & \vert z_3\vert \\
   \frac{14}{17} & -\frac{112}{199} & \vert z_4\vert \\
   \frac{18}{97} & -\frac{44}{51} & \vert z_5\vert \\
   -\frac{45}{76} & -\frac{71}{134} & \vert z_6\vert \\
   -\frac{32}{119} & \frac{2}{33} &-\vert z_7\vert \\
    \frac{72}{85} & \frac{50}{113} & \vert z_8\vert \emph{\emph{}}\\
 \end{pmatrix},
\end{eqnarray}
where the $k$-th row of the above matrix is understood to be $\vec{x}_{k}$, and $|z_k|=\sqrt{1-v_{k1}^2-v_{k2}^2}$.

Similarly, the 9-setting steering inequality is given by
\begin{align}\label{9inequality}
\sum^{9}_{i=1}\sum^{3}_{j=1}s_{ij}\langle ab\rangle_{ij}+\sum^{9}_{i=1}s^{A}_{i}\langle a\rangle_{i}+\sum^{3}_{j=1}s^{B}_{j}\langle b\rangle_{j}\leq L,
\end{align}
with $L=1$, and
\begin{eqnarray}
&& \bf{S} = \left( \begin{array}{ccc}
\frac{53}{521} & \frac{53}{902} & -\frac{47}{271}\\
   -\frac{9}{280} & -\frac{13}{379} & -\frac{44}{273}\\
  -\frac{39}{220} & -\frac{53}{419} & -\frac{17}{312}\\
   \frac{34}{471} & -\frac{16}{339} & \frac{8}{55}\\
   \frac{115}{2184} & -\frac{13}{426} & -\frac{63}{685}\\
   \frac{23}{404} & -\frac{43}{354} & -\frac{34}{285}\\
   \frac{29}{185} & -\frac{130}{873} & \frac{15}{289}\\
   -\frac{26}{147} & \frac{2}{387} & -\frac{2}{111}\\
   \frac{7}{132} & \frac{110}{353} & -\frac{21}{358}\end{array} \right),
 \bf{S^{A}}= \left( \begin{array}{c}
 -\frac{75}{974} \\
   -\frac{23}{325} \\
  -\frac{7}{305} \\
   \frac{25}{389}\\
   -\frac{11}{272} \\
   -\frac{39}{751} \\
   \frac{11}{467}\\
   -\frac{1}{131}\\
   -\frac{10}{359}\end{array} \right),
 \bf{S^{B}}=\left( \begin{array}{c}
 \frac{1}{564} \\
   \frac{1}{161} \\
  -\frac{26}{59} \end{array} \right).
  \end{eqnarray}
One can show that the quantum violation for the state $\rho_{AB}(\alpha=1/2)$ is $\frac{119}{116}\backsimeq 1.0258$, where the 9 measurement settings of Alice can be characterized by Bloch vectors $\vec{x}_{i}$ ($i=1, 2,...,9$), which are
\begin{eqnarray}
\bf{V}=\begin{pmatrix}
-\frac{272}{453} & -\frac{43}{124} & \vert z_1\vert \\
   \frac{43}{165} & \frac{32}{115} & \vert z_2\vert \\
  \frac{129}{161} & \frac{251}{439} & \vert z_3\vert \\
   -\frac{7}{13} & \frac{25}{71} & -\vert z_4\vert \\
   -\frac{68}{115} & \frac{37}{108} & \vert z_5\vert \\
   -\frac{47}{131} & \frac{49}{64} & \vert z_6\vert \\
   -\frac{123}{172} & \frac{36}{53} &-\vert z_7\vert \\
   \frac{314}{315} & -\frac{3}{103} & \vert z_8\vert \\
   -\frac{90}{541} & -\frac{89}{91} & \vert z_9\vert \emph{\emph{}}\\
 \end{pmatrix}.
\end{eqnarray}

\emph{Remark 4.---} If the parameter $\alpha$ is not required to run over all the region of $[0,1]$, one may also present other class of more efficient one-way steerable states. For the the simplest case, one may just select
\begin{eqnarray}\label{f0}
f=0,
\end{eqnarray}
i.e., the parameter is independent of $\alpha$. Correspondingly from Eq. (\ref{state-2c}) one has the state as
\begin{eqnarray}\label{state-2c1}
\rho_{AB}(\alpha) = \frac{1}{4} \biggr(\openone\otimes \openone +\frac{1}{3} \;\sigma_{3} \otimes \openone -\frac{1}{2}\; \openone\otimes \sigma_3
-\alpha\sum_{k=1}^3  \sigma_k
\otimes \sigma_k \biggr).
\end{eqnarray}
However, the matrix is a density matrix only if $\alpha\in [0, \frac{1}{18}(6+\sqrt{69})]$. For $\alpha=1/2$, the state (\ref{state-2c1}) is identical to the state (\ref{state-2d}), they all violate the 8-setting and the 9-setting steering inequality.

Eventually, we would like to mention that, due to the difficulty in numerical computations we are not able to obtain the optimal states for demonstrating the one-way EPR steering, which is a difficult problem. However, based on our result, the amount of quantum violation is much larger but achieved with fewer measurements in comparison to the previous result in \cite{Bowles}. To our knowledge, we do not know whether the quantum violation by inequality (\ref{9inequality}) could be observed with current quantum technology. However, we believe that this result would be interesting and helpful for the experimenters.